\documentclass[aps,pra,graphicx,9pt,twocolumn,superscriptaddress]{revtex4-2}
\usepackage[colorlinks,linkcolor=blue,urlcolor=blue,anchorcolor=blue,citecolor=blue]{hyperref}
\usepackage{amsmath}
\usepackage{graphicx}
\usepackage{dcolumn}
\usepackage{mathrsfs}
\usepackage{physics}
\usepackage{amssymb}
\usepackage{amsfonts,multirow}
\usepackage{bm}
\usepackage{changes}
\definechangesauthor[name={HLZhang}, color=red]{HL}
\UseRawInputEncoding
\begin{document}
\title{Critical quantum geometric tensors of parametrically-driven nonlinear resonators}
\author{Hao-Long Zhang}
\author{Jia-Hao L${\rm \ddot{u}}$}
\author{Ken Chen}\author{Xue-Jia Yu}
\email{xuejiayu@fzu.edu.cn}
\author{Fan Wu}
\email{t21060@fzu.edu.cn}
\author{Zhen-Biao Yang}
\email{zbyang@fzu.edu.cn}

\author{Shi-Biao Zheng}

\address{Fujian Key Laboratory of Quantum Information and Quantum Optics, College
    of Physics and Information Engineering, Fuzhou University, Fuzhou, Fujian 350108, China}

\begin{abstract}

    Parametrically driven nonlinear resonators represent a building block for realizing fault-tolerant quantum computation and are useful for critical quantum sensing. From a fundamental viewpoint, the most intriguing feature of such a system is perhaps the critical phenomena, which can occur without interaction with any other quantum system. The non-analytic behaviors of its eigenspectrum have been substantially investigated, but those associated with the ground state wavefunction have largely remained unexplored. Using the quantum ground state geometric tensor as an indicator, we comprehensively establish a phase diagram involving the driving parameter $\varepsilon$ and phase $\phi$. The results reveal that with the increase in $\varepsilon$, the system undergoes a quantum phase transition from the normal to the superradiant phase, with the critical point unaffected by $\phi$. Furthermore, the critical exponent and scaling dimension are obtained by an exact numerical method, which is consistent with previous works. Our numerical results show that the phase transition falls within the universality class of the quantum Rabi model. This work reveals that the quantum metric and Berry curvature display diverging behaviors across the quantum phase transition.
\end{abstract}

\maketitle

\section{Introduction}
    Physical systems can change between two distinct phases, exemplified by the ice-water transformation. Classical phase transitions are incurred by thermal fluctuations, while the quantum counterparts are triggered by quantum fluctuations, even at zero temperature, giving rise to quantum phase transitions (QPTs) \cite{RN14, RN15,cardy1996scaling,sachdev2023quantum}. There exist different types of QPTs \cite{RN16,xu2012unconventional,yu2022prb,yu2022prl,yang2023emergent}, among which of special interest are equilibrium ones. These phase transitions are manifested by the non-analytic behaviors of the ground state of the Hamiltonian at a critical point, around which a tiny change of the driving parameter would lead to a dramatic response. Such consequences are referred to as the critical phenomena \cite{RN17,coleman2005quantum}, which usually occurs in the thermodynamic limit.

    Conventionally, the system size regarding thermodynamic limit refers to the number of the interacting constituents in the entire system, e.g., the number of spins that are coupled to a bosonic mode in the Dicke model \cite{RN21, RN22,RN18}, which can exhibit a superradiant phase transition at the critical point where the spin-boson coupling strength is comparable to the system frequencies. One decade ago, it was realized that the superradiant phase transition can also occur in the Rabi model \cite{RN19, RN20}, which comprises only a single spin coupled to a bosonic mode. With this model, the thermodynamic limit is replaced by the scaling limit, where the ratio between the frequency of the qubit and the bosonic mode tends to infinity. The past few years have seen several impressive experimental demonstrations of such phase transitions \cite{RN21, RN22,RN23, RN25}.

    More recently, QPTs were revealed with an even simpler system, which involves a single resonator, featuring a competition between the Kerr nonlinearity and parametric drive \cite{RN26, arxiv:2310.13636, PRA.102.6.063531,npj.9.1.76,PhysRevA.94.033841,PhysRevA.103.033711}. Such a system can be employed to construct a Kerr cat qubit with biased noises that are useful for realizing fault-tolerant quantum computation \cite{RN27, RN28, RN29, RN30, RN31, RN32, RN33,PRXQuantum.4.020337}, quantum transducer \cite{PhysRevLett.123.173601} and offers a possibility for criticality-enhanced quantum sensing \cite{RN34, RN35}. Physically, the infinity of the dimension of the Hilbert space plays the role of the thermodynamic limit. In such a space, the photons interact with each other through the Kerr effect. This interaction, together with the driving, can lead to a superradiant phase transition in the equilibrium state, like the Dicke or Rabi model, but without the participation of any spin. The associated critical phenomena are useful in sensitivity enhancement. The previous investigations of the phase transition in such a system primarily focused on the non-analytic behaviors of the ground state energy and the excitation energy \cite{RN26}, leaving the geometric aspects of wavefunction properties across parameter variations largely unexplored.

    In this work, we explore the critical phenomena associated with the ground state wavefunction across the equilibrium phase transition of a single parametrically-driven Kerr resonator. Specifically, we quantify non-analytic behaviors using the quantum geometric tensor (QGT) \cite{RN39, PhysRevB.103.174104}, consisting of the quantum metric \cite{RN37, RN38, OptExp.25.31.41669,PhysRevB.106.165124,PhysRevE.107.054122} and the Berry curvature \cite{RN41, PhysRevLett.96.077206,hamma2006berry}. Employing an exact numerical diagonalization method, we systematically delineate a phase diagram with the driving parameter $\varepsilon$ and phase $\phi$. The results reveal that a QPT from the normal phase to the superradiant phase occurs as $\varepsilon$ increases, with the critical point unaffected by $\phi$. Additionally, the critical exponent and scaling dimension are determined using finite-size scaling analysis, consistent with previous results \cite{PhysRevA.97.013845, RN20}. Our findings demonstrate that the phase transition falls within the universality class of the quantum Rabi model.

The rest of this paper is organized as follows: Sec.~\ref{sec2} contains the model of a single parametrically-driven Kerr resonator, the numerical method employed, and the global phase diagram. Sec.~\ref{sec3} shows the scaling relations of QGT. Sec.~\ref{sec4} presents the finite-size scaling of critical behavior, followed by a brief comparison with the two-photon driving bosonic model without Kerr nonlinearity. The conclusion
is presented in Sec.~\ref{sec5}. Additional data for our numerical calculations are provided in the Appendixes.

    \begin{figure} 
        \centering
        \includegraphics{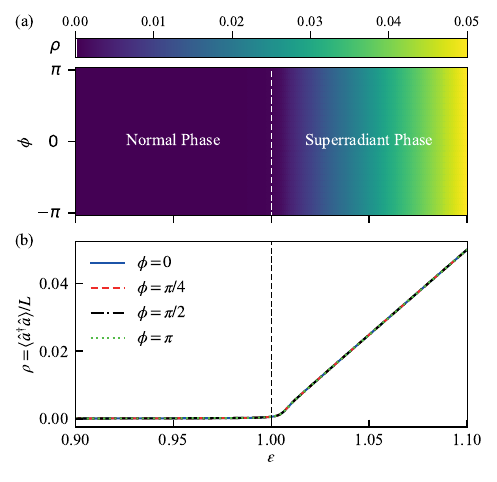}
        \caption{(a) The ground-state phase diagram of parametrically-driven nonlinear resonator with respect to control parameter $\varepsilon$ and $\phi$ characterized by the rescaled average photon number $\rho=\langle a^{\dagger} a\rangle/L$, which plays the role of order parameter in the description of phase transition. The phase boundary is close to $\varepsilon=1$, which coincides with the theoretical expectation. (b) The rescaled average photon number as a function of $\varepsilon$ with $L=10^4$ and different $\phi=0,\pi/4,\pi/2,\pi$. All curves coincide together, indicating that the process of phase transition is independent of $\phi$.}
        \label{fig1}
    \end{figure}
\section{ Quantum phase diagram of parametrically-driven nonlinear resonator}
\label{sec2}
    \subsection{Model and method}
      The system under investigation is a parametrically-driven Kerr resonator, described by the following Hamiltonian ($\hbar = 1$ is set)
        \begin{equation}
        \label{model}
             H_{\rm{driven}} = \delta  a^{\dagger} a- P a^{\dagger 2}-P^* a^2 + K a^{\dagger 2} a^2,
        \end{equation}
        where $ a$ and $ a^{\dagger}$ represent the photon annihilation and creation operators, respectively, with the canonical commutation relation $[ a, a^{\dagger}]=1$. The Kerr nonlinearity is denoted by $K$, and $P$ is the amplitude of the two-photon drive with detuning $\delta$. Notably, this Hamiltonian exhibits a discrete $\mathbb{Z}_{2}$ symmetry, being invariant under the parity transformation $U= e^{i\pi  a^{\dagger} a}$. To analyze how the QPT is influenced by a finite value of $\delta/K$, we perform a finite-size scaling analysis, based on the effective ``system size" defined as $L = \delta/K$, with $L$ governing the proximity to the thermodynamic limit. To investigate the geometric properties of the model, we can parametrize $P = \delta\varepsilon \exp(-i\phi)/2$, and study QPTs in terms of two-photon drive amplitude $\varepsilon$ and phase $\phi$. Then the Hamiltonian can be rewritten as 
        \begin{equation}\label{Hd}
            {\rm{H}}^{\prime}_{\rm{driven}} =K  a^{\dagger 2} a^2 + \delta a^{\dagger} a  - \frac{\delta\varepsilon}{2}(e^{-i\phi} a^{\dagger 2} + e^{i\phi} a^2).
        \end{equation}
        Before delving into QPTs, it is pertinent to discuss the existing phases in the model. As a first step, we can analytically obtain quantum phases under different limits when $K\rightarrow 0$. On the one hand, for $\varepsilon < 1$, the first term can be neglected, and a squeezing transformation $S(r_{\rm n})=e^{(r_{\rm n}^* a^2-r_{\rm n} a^{\dagger 2})/2}$ is applied, with $r_{\rm n}=\frac{1}{4}\ln \frac{1-\varepsilon}{1+\varepsilon}e^{-i\phi}$. This transforms the Hamiltonian to the normal phase Hamiltonian:
        \begin{equation}
                {\rm{H}}_{\rm n}=\omega_{\rm n}^{e} a^{\dagger} a+\omega_{\rm n}^{g},
        \end{equation}
        where $\omega_{\rm n}^{e}=\delta\sqrt{1-\varepsilon^2}$ and $\omega_{\rm n}^{g}=\delta(\sqrt{1-\varepsilon^2}-1)/2$ are the excitation and ground-state energies in the normal phase, respectively. Therefore, the system ground state is given by:  
        \begin{equation}\label{psign}
            \ket{\psi_{\rm n}^{g}}=S(r_{\rm n})\ket{0}.
        \end{equation}
        It is worth noting that this wavefunction is a squeezed vacuum state, which preserves the $\mathbb{Z}_{2}$ symmetry. The squeezing parameter is dependent on $\varepsilon$ and $\phi$.
        
        The energy gap vanishes and the squeezing parameter tends to infinity at the critical point $\varepsilon_c=1$. To reveal the ground state in the regime $\varepsilon>1$, a displacement transformation $D(\pm\alpha)$ is performed, with $\alpha=\sqrt{L(\varepsilon-1)/2}\exp(-i\phi/2)$. Discarding higher-order terms due to the large $L$, the original Hamiltonian becomes:
        \begin{equation}\label{HD'}
             H_{\rm{D}}\approx \delta a^\dagger a-\Omega(e^{i\phi} a^2+e^{-i\phi} a^{\dagger 2})-L\delta^2(\varepsilon-1)^2/4,
        \end{equation}
        where $\delta=\delta(2\varepsilon-1)$ and $\Omega=\delta/2$. Then we perform the squeezing transformation $S(r_{\rm s})=e^{(r_{\rm s}^* a^2-r_{\rm s} a^{\dagger 2})/2}$ and set $r_{\rm s} = \frac{1}{4}\ln\frac{\varepsilon-1}{\varepsilon}e^{-i\phi}$, transforming the Hamiltonian to the superradiant phase Hamiltonian: 
        \begin{equation}
             H_{\rm s} =\omega_{\rm s}^{e}  a^\dagger a+\omega_{\rm s}^{g},
        \end{equation}
        where $\omega_{\rm s}^{e}=2\delta\sqrt{\varepsilon(\varepsilon-1)}$ and $\omega_{\rm s}^{g}=\delta(\sqrt{\varepsilon(\varepsilon-1)}-\varepsilon+1/2)-L\delta^2(\varepsilon-1)^2/4$. Consequently, the system has two degenerate ground states, given by
        \begin{equation}
            \ket{\psi_{\rm s(\pm)}^{g}}=D(\pm\alpha)S(r_{\rm{s}})\ket{0},
        \end{equation}
        both breaking the $\mathbb{Z}_{2}$ parity symmetry.
        
        The above results indicate that the system undergoes a transition from the normal phase to the superradiant phase at the critical point $\varepsilon_c=1$. In general, when $K$ takes a finite value, obtaining the ground state analytically becomes challenging. However, we can numerically diagonalize the Hamiltonian in the representation of the Fock basis and obtain the corresponding eigenstates by solving the secular equation. This method requires the photon number cut-off much larger than the average photon number of the system so that all the information on the quantum state can be included. In this work, unless explicitly stated otherwise, we set the cut-off to $N_{\text{cut}}=800$.
        
    \subsection{Quantum phase diagram}
     To explore the global phase diagram of the model, we numerically calculate the average photon density $\rho=\langle a^{\dagger} a\rangle/L$ as a function of the two-photon drive amplitude $\varepsilon$ and phase $\phi$. This average photon density serves as the order parameter in the theory of QPT. As depicted in Fig.~\ref{fig1}(a), for a specific $\phi$, when $\varepsilon \textless 1 $, the average photon density remains at 0, indicating the stability of the normal phase even with finite $K$. Conversely, when $\varepsilon \textgreater 1$, the average photon density abruptly increases [see Fig.~\ref{fig1}(b)], signifying that, photons spontaneously condense, breaking the $\mathbb{Z}_{2}$ parity symmetry and exhibiting the superradiant phase. Moreover, it is observed that the normal-to-superradiant phase transition point is unaffected by the two-photon driving phase $\phi$, which will be discussed in the context of the QGT below.

    Having established the global phase diagram, the subsequent crucial task is to investigate the QPT between different phases and determine the universality class to which it belongs. Various methods exist for detecting quantum critical points and determining their universality class\cite{2018Lectures,cardy1996scaling}. In the following sections, we introduce the basic concept of the critical QGT and employ it to examine QPT within the phase diagram.
    
\section{Basic notion for critical QGT}
\label{sec3}
The system undergoes a continuous phase transition from an ordered to a disordered phase when tuning the external field $\lambda$ to a critical value $\lambda_{c}^{*}$, at which the structure of the ground-state wavefunction changes significantly. More generally, if we consider a non-degenerate quantum system $ H(\bm{\lambda})$ that depends smoothly on a set of $M$ real adiabatic parameters denoted as $\bm{\lambda} = (\lambda_1, \lambda_2, ..., \lambda_M)$, for the ground-state eigenvector $\ket{u_{0}(\bm{\lambda})}$ with the energy $E_{0}$, the QGT is defined by

\begin{equation}\label{Qmunu}
Q_{jk}=\bra{\partial_j u_{\rm 0}(\bm{\lambda})}(1-\ket{u_{\rm 0}(\bm{\lambda})}\bra{u_{\rm 0}(\bm{\lambda})})\ket{\partial_k u_{\rm 0}(\bm{\lambda})},
\end{equation}

where $\partial_j=\partial/\partial{\lambda_{j}}$. The (symmetric) real part of the QGT yields the quantum metric tensor $g_{jk}=\Re{Q_{jk}}$ \cite{RN37, RN38, OptExp.25.31.41669, photonics10030256}, which is a Riemannian metric providing the distance $ds^{2} = g_{jk}d\lambda_{j} d\lambda_{k}$ between the quantum states $\ket{u_{0}(\lambda)}$ and $\ket{u_{0}(\lambda+d\lambda)}$, corresponding to infinitesimally different parameters. The (antisymmetric) imaginary part of the QGT encodes the Berry curvature $\mathcal{F}_{jk}=-2\Im{Q_{jk}}$ \cite{RN41, qute.202300068, PhysRevD.95.046010}, which, when integrated over a surface subtended by a closed path in the parameter space, gives rise to the geometric Berry phase.

\begin{figure}
\centering
\includegraphics{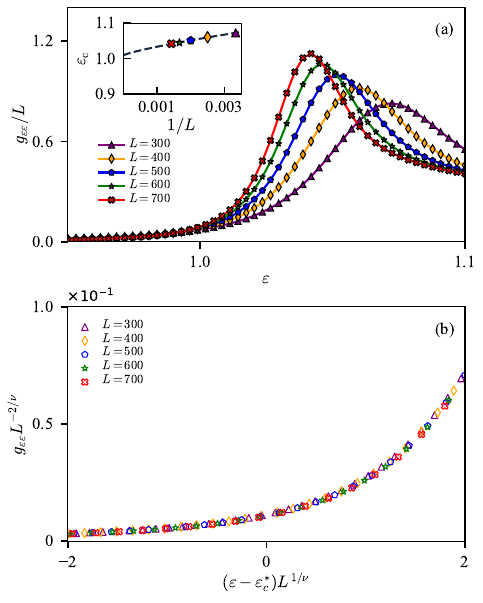}
\caption{(a) The rescaled quantum metric $g_{\varepsilon\varepsilon}/L$ for $L=300, 400, 500, 600, 700$ sites as a function of the control parameter $\varepsilon$. The inset is the extrapolation of the critical point $\varepsilon_{\rm c}$, symbols denote the finite-size result for $L=300, 400, 500, 600, 700$ sites. We use polynomial fitting $\varepsilon_{\rm c}(L) = \varepsilon^{*}_{\rm c} + aL^{-b}$ and extrapolate the critical point $\varepsilon^{*}_{\rm c}\approx 1.008$. (b) Data collapse of $g_{\varepsilon\varepsilon}L^{-2/\nu}$ as a function of $L^{1/\nu}(\varepsilon-\varepsilon_c^{*})$. The results show the quantum metric $g_{\varepsilon\varepsilon}$ for different $L$ collapsing perfectly only when $\nu=1.510$.}
\label{fig2}
\end{figure}

An important aspect of the QGT is that its singularities are associated with QPTs. To gain a better understanding of this aspect, we rewrite Eq.~(\ref{Qmunu}) as

\begin{equation}\label{QmunuH}
Q_{jk}=\sum_{{\rm n} \neq 0}\frac{\bra{u_{\rm n}}\partial_j  H \ket{u_{\rm 0}}\bra{u_{\rm 0}}\partial_k  H \ket{u_{\rm n}}}{(E_{\rm n}-E_{\rm 0})^2},
\end{equation}
where $\ket{u_{\rm n}}$ is the eigenstate with eigenenergy $E_{\rm n}$. This expression clearly suggests that at the critical points, where the energy gap between the ground state and the nearest excited state vanishes in the thermodynamic limit, the QGT might show a singular behavior. If there is a local perturbation operator in the Hamiltonian, then under the scaling transformation \cite{RN47}, in the vicinity of the critical point $\lambda_{c}^{*}$, both the perturbation operator and the rescaled QGT $Q_{jk}/L^d$ have well-defined scaling dimensions $\Delta_{\mu} $ ($\mu = \varepsilon, \phi$ in our case, see the next section) and $\Delta_{jk} = \Delta_j + \Delta_k - 2z-d$, respectively, where $z$ is the dynamical exponent, and $d$ is spatial dimension of the system.

For a continuous QPT of a finite system with size $L$, the QGT exhibits a peak at the pseudo-critical point $\lambda_c(L)$, and the value of the quantum critical point $\lambda_{c}^{*}$ can be estimated by polynomial fitting $\lambda_{c}(L)=\lambda_{c}^{*}+aL^{-b}$~\cite{10.1063/1.3518900}. In the vicinity of $\lambda_{c}(L) = \lambda_{c}$, previous studies \cite{RN47} have shown that the finite-size scaling behaviors of the QGT follow

\begin{equation}\label{Qjkgamma}
Q_{jk}(\lambda=\lambda_{\rm c})\propto L^{\Delta_{jk}},
\end{equation}
and~\cite{RN45,konig2016prb,gu2014spectral}

\begin{equation}\label{data}
\abs{Q_{j k}}L^{-\Delta_{j k}}=f_{\text{QGT}}((\lambda-\lambda_{\rm c}^{*})L^{1/\nu}),
\end{equation}
where $L$ is the effective system size, $\nu$ is the correlation length exponent, and $f_{\text{QGT}}$ is an unknown scaling function.

Based on Eq.~(\ref{Qjkgamma}) and (\ref{data}), the values of critical exponents $\nu$ and the scaling dimension of different perturbations can be determined. However, as shown in the preceding section, the driving parameter for the QPT is $\varepsilon$ instead of $\phi$. In this work, we mainly focus on the scaling behavior of $Q_{\varepsilon\varepsilon}$, $Q_{\varepsilon\phi}$, and $Q_{\varepsilon \phi}$. 

    \begin{figure}
        \flushleft
        \includegraphics[width=0.48\textwidth]{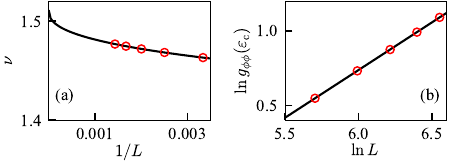}
        \caption{(a) The correlation length exponent as the function of $1/L$ extracted by fitting the slope of $\ln(g_{\varepsilon\varepsilon})$ in the quasi-critical point as a function of $\ln(L)$. The hollow circles represent the numerical results and the black solid line donates the function $y=ax^b+c$, fitted with the numerical data. As $L$ increases, the correlation length exponent $\nu$ converges to $1.510$. (b) The finite size scaling of $g_{\phi\phi}$ at a critical point as a function of $\ln(L)$. The hollow circles are the exact numerical results and the black solid line is the linear fit of the numerical data with the fitting function $y=ax+b$. The fit scaling dimension is $\Delta_{\phi\phi}\approx0.643$.}
        \label{fig3}
    \end{figure} 
  
\section{Finite size scaling and critical exponents}
\label{sec4}
\subsection{Quantum critical behavior with Kerr nonlinearity}
\label{seciv4}
    The next question is what is the critical behavior of the parametrically-driven nonlinear resonator with finite Kerr nonlinearity and to which universality class it belongs. To this end, we perform a numerical diagonalization of the Hamiltonian in the Fock basis and subsequently evaluate the components of the QGT using Eq. (\ref{QmunuH}). The numerical results depend on the dimensions of the Hilbert space, and convergence is observed as the photon number cut-off increases. We extract the convergent results as a function of the driving parameter $\varepsilon$ for different $L$. Subsequently, we conduct fits to obtain the corresponding scaling dimension $\Delta_{jk}$ and correlation length exponent $\nu$.
    
    An interesting finding is that the first diagonal component of the QGT, denoted as $g_{\varepsilon\varepsilon}$, is equivalent to the fidelity susceptibility \cite{RN43, RN44, PhysRevB.106.165124, PhysRevE.107.054122, RN45, RN46,gu2009scaling,RN46,wang2015prx,RN44} (see Appendix~\ref{appendixA}). The corresponding scaling dimension is related to the correlation length exponent as $\Delta_{\varepsilon\varepsilon}=2/\nu$. Illustrated in Fig.~\ref{fig2}(a), the rescaled quantum metric tensor $g_{\varepsilon\varepsilon}/L$ exhibits a peak that becomes sharper with increasing size at the pseudo-critical point $\varepsilon_{c}(L)$. This implies that $L$ plays a role analogous to the system size in an ordinary phase transition. In the inset of Fig.~\ref{fig2}(a), we perform finite-size scaling of the pseudo-critical point $\varepsilon_{c}(L)$ as a function of inverse effective system sizes $1/L$. The quantum critical point $\varepsilon_{c}^{*}$ is estimated by polynomial fitting: $\varepsilon_{c}(L) = \varepsilon_{c}^{*} + a L^{-b}$. Consequently, at the critical point $\varepsilon_{c}$, the corresponding scaling dimension $\Delta_{\varepsilon \varepsilon}$ is obtained by fitting the function in Eq.(\ref{Qjkgamma}), and the correlation length exponent $\nu$ is calculated. As shown in Fig.~\ref{fig3}(a), we find that the scaling dimension $\Delta_{\varepsilon\varepsilon}\approx1.325$ and correlation length exponent $\nu$ converges to $1.510$ as the thermodynamic limit is reached, closely matching the result $1.5$ for quantum Rabi models in the infinite qubit-field frequency ratio limit \cite{RN20}. 
       \begin{figure}
        \centering
        \includegraphics{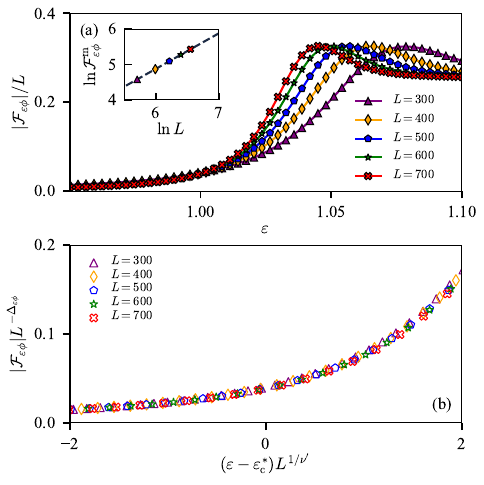}
        \caption{(a) The rescaled Berry curvature $\mathcal{F}_{\varepsilon\phi}$ as a function of control parameter $\varepsilon$. The inset displays the finite size scaling of $\mathcal{F}_{\varepsilon\phi}$ at the quasi-critical point as a function of $\ln(L)$. The solid circles denote the exact numerical results and the black dashed line is the linear fit, which yields a scaling dimension $\Delta_{\varepsilon\phi}\approx 1.0$. (b) Data collapse of the $|\mathcal{F}_{\varepsilon\phi}|L^{-\Delta_{\varepsilon\phi}}$, which is a function of $L^{1/\nu^{\prime}}(\varepsilon-\varepsilon_c^{*})$ only with $\nu^{\prime}=1.5$ so that for different $L$ all the lines collapse perfectly.}
        \label{fig4}
    \end{figure}
    
    To verify the universal scaling behavior of the QGT, according to Eq.(\ref{data}), the tensor can be scaled by $L^{-\Delta_{jk}}g_{\varepsilon \varepsilon}$ as a function of $L^{1/\nu} |\varepsilon - \varepsilon_{c}^{*}|$ in the vicinity of the quantum critical point $\varepsilon_{c}^{*}$. Upon inserting the obtained critical point $\varepsilon_{c}^{*}$ and correlation length exponent $\nu$ into Eq.(\ref{data}), all metric tensors for different $L$ collapse into a single one [Fig.~\ref{fig2}(b)], indicating that $g_{\varepsilon \varepsilon}$ is a universal function of $L^{1/\nu} |\varepsilon - \varepsilon_{c}^{*}|$, and the estimated critical point and critical exponent $\nu \approx 1.510$ are accurate.
    \begin{figure}
        \centering
        \includegraphics[width=0.47\textwidth]{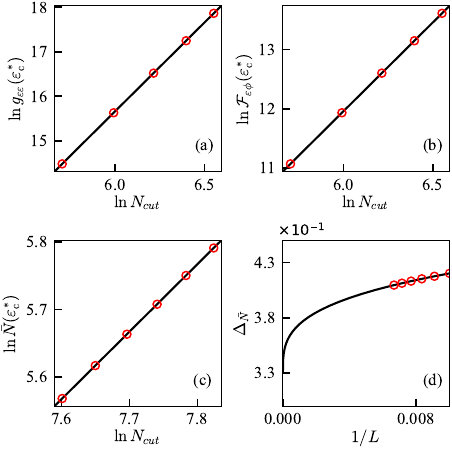}
        \caption{(a) Finite size scaling of $g_{\varepsilon\varepsilon}$ at critical point with respect to the photon cut-off $N_{cut}$ without nonlinearity $K=0$. The fitted critical exponent is $\gamma_{1}\approx 3.996$. (b) Finite-size scaling of $\mathcal{F}_{\varepsilon\phi}$ at critical point with respect to the photon cut-off $N_{cut}$ without nonlinearity $K=0$. The fitted scaling dimension is $\gamma_{2}\approx 2.997$. (c) Finite-size scaling of average photon number $\bar{N}=\langle a^{\dagger} a\rangle$ for the photon number cut-off $N_{cut}$ without nonlinearity. The fitted scaling dimension approximates $\alpha\approx0.998$. (d) The critical exponent of the average photon number as a function $1/L$. The fitted scaling dimension $\Delta_{\bar{N}}$ converges to $0.330$. The hollow circles are the exact numerical results, and the black solid line is the linear fit of the numerical data.}
        \label{fig5}
    \end{figure}
    
    Furthermore, we observe that the second diagonal component of the QGT $g_{\phi\phi}$ also diverges at the critical point $\varepsilon_{c}^{*}$ due to gaplessness in the limit $L\rightarrow \infty$. As shown in Fig.~\ref{fig3}(b), the values of $g_{\phi\phi}$ at the critical point diverge as $L$ increases, with a scaling dimension $\Delta_{\phi\phi}\approx 0.643$.
    
    Finally, we delve into the imaginary part of the off-diagonal component of the QGT, denoted as the Berry curvature $\mathcal{F}_{\varepsilon \phi}$. As illustrated in Fig.~\ref{fig4}(a), the corresponding scaling dimension is $\Delta_{\varepsilon\phi}\approx 1.0$. We obtatin the correlation length exponent and critical point by fitting the metric tensor into Eq.~(\ref{data}). Similarity, all Berry curvatures for different $L$ collapse into a single one in Fig.~\ref{fig4}(b), which also suggests $ \mathcal{F}_{\varepsilon \phi}$ is a universal function of $L^{1/\nu} |\varepsilon - \varepsilon_{c}^{*}|$. 
    
    To summarize, our numerical findings collectively affirm that the critical point of the parametrically-driven Kerr resonator aligns with the same universality class as the Rabi model \cite{RN20}.

    As an additional note, combined with the previous scaling analysis, we have determined the scaling dimensions $\Delta_{jk}$ of the geometric tensor $Q_{jk}$. Referring to Eq.(6) in Ref. \cite{RN47}, we can derive the scaling dimensions for $\varepsilon$ and $\phi$ perturbations, respectively. The results reveal that $\Delta_{\varepsilon} = 0.3375$ and $\Delta_{\phi} = 0.6785$. Additionally, the scaling dimension of the Berry curvature is $ 2-\Delta_{\varepsilon} - \Delta_{\phi} = 0.984$, closely matching the result obtained from the finite-size scaling $\Delta_{\varepsilon \phi} \approx 1.0$. Crucially, as $\Delta_{\varepsilon} \textless \Delta_{\phi}$, this implies that in the renormalization group sense, $\varepsilon$ perturbation is more relevant than $\phi$ perturbation. In other words, the driving parameter for the QPT is the $\varepsilon$ perturbation, consistent with the findings illustrated in our phase diagram.
    
\subsection{Quantum critical behavior without Kerr nonlinearity}
    
When $K=0$, the model reduces to a two-photon driven bosonic mode \cite{RN35}. A natural question is whether the inclusion of finite Kerr nonlinearity alters the universality class of the phase transitions. For this purpose, we specifically examine the case without Kerr nonlinearity. Diverging from the previous situation, the photon number cut-off $N_{{\rm{cut}}}$ serves as the effective system size to quantify the approach to the thermodynamic limit. At this time, the finite-size scaling relation of the first diagonal component of the QGT at the quantum critical point $\varepsilon^{*}_{c}=1$ can be written as $g_{\varepsilon\varepsilon}(\varepsilon^{*}_{\rm c})\propto N_{{\rm{cut}}}^{\gamma_1}$, where $\gamma_1$ signifies the scaling dimension of $g_{\varepsilon\varepsilon}$ concerning the photon number cut-off. The numerical result is depicted in Fig.~\ref{fig5}(a), revealing a fitted scaling dimension of approximately $\gamma_1 \approx 3.996$. Similarly, the finite-size scaling relation of the Berry curvature is given by $\mathcal{F}_{\varepsilon\phi}(\varepsilon^{*}_{\rm c}) \propto N_{cut}^{\gamma_2}$, where $\gamma_2$ represents the scaling dimension of $\mathcal{F}_{\varepsilon\phi}$ with respect to the photon number cut-off. As illustrated in Fig.~\ref{fig5}(b), the fitted scaling dimension is approximately $\gamma_2\approx 2.997$. For easier comparison, we write $g_{\varepsilon \varepsilon}$ and $\mathcal{F}_{\varepsilon\phi}$ as scaling relations for the average number of photons $\bar{N}$ for both $K \neq$ and $K=0$. Intuitively, when the photon number cutoff $N_{{\rm{cut}}}$ is increased, the average photon number of the system $\bar{N}=\langle a^{\dagger} a \rangle $ becomes larger, that is, $\bar{N} \propto N_{{\rm{cut}}}^{\alpha}$, as shown in the Fig.~\ref{fig5}(c), where we find $\alpha \approx 0.998$. Then the finite-size relation in terms of average photon number $\bar{N}$ can be expressed as ($K = 0$)
\begin{equation}
\label{eqk}
\begin{split}
    &g_{\varepsilon\varepsilon}(\varepsilon^{*}_{\rm c})\propto \bar{N}^{\alpha \gamma_1} = \bar{N}^{\beta_1},\\
    &\mathcal{F}_{\varepsilon\phi}(\varepsilon^{*}_{\rm c})\propto \bar{N}^{\alpha \gamma_2} = \bar{N}^{\beta_2}.
\end{split}
\end{equation}

On the other hand, in the previous section \ref{seciv4}, the scaling relation of $g_{\varepsilon\varepsilon}$ and $\mathcal{F}_{\varepsilon\phi}$ concerning effective system size $L=\delta/K$ are expressed as $g_{\varepsilon \varepsilon}\propto L^{\Delta_{\varepsilon\varepsilon}}$ and  $\mathcal{F}_{\varepsilon \phi}\propto L^{\Delta_{\varepsilon\phi}}$, respectively. Moreover, the scaling relation between the average number of photons and the effective system size is $\bar{N}(\varepsilon^{*}_{\rm c})\propto L^ {\Delta_{\bar{N}}}$. As shown in Fig.~\ref{fig5}(d), $\Delta_{\bar{N}} \approx 0.33$. Combining the aforementioned scaling formulas, the finite-size scaling of $g_{\varepsilon \varepsilon}$ concerning the average photon number is determined as ($K \neq 0$)

\begin{equation}
\label{eqk1}
\begin{split}
    &g_{\varepsilon\varepsilon}(\varepsilon^{*}_c)\propto \bar{N}^{\Delta_{\varepsilon \varepsilon}/\Delta_{\bar{N}}} = \bar{N}^{\beta_1^ {\prime}},\\
    &\mathcal{F}_{\varepsilon\phi}(\varepsilon^{*}_c)\propto \bar{N}^{\Delta_{\varepsilon \phi}/\Delta_{\bar{N}}} = \bar{N}^{\beta_2^ {\prime}}.
\end{split}
\end{equation}

Here $\beta_{1}^{\prime}$ and $\beta_{2}^{\prime}$ indicate the scaling dimension of $g_{\varepsilon\varepsilon}$ and $\mathcal{F}_{\varepsilon \phi}$ concerning the average photon number, respectively. With all the numerical results obtained, we ascertain the scaling dimension of the quantum metric tensor concerning the average number of photons $\beta_1^{\prime} = \Delta_{\varepsilon \varepsilon}/\Delta_{\bar{N}} \approx 4.04$, which is approximately equal to $\beta_1 = \alpha \gamma_1 \approx 3.99$. Similarly, the Berry curvature with respect to the average photon number yields $\beta_2^{\prime} = \Delta_{\varepsilon \phi}/\Delta_{\bar{N}} \approx 3.03$, roughly equal to $\beta_2=\alpha \gamma_2 \approx 2.99$. These results indicate that finite Kerr nonlinearity does not alter quantum critical behavior and always belongs to the same universality class as the quantum Rabi model.

\section{Conclusion} 
\label{sec5}
    In summary, we have studied the properties of the ground state wavefunction and critical behavior of the parametrical-driven Kerr resonator through numerical simulations. Specifically, using critical QGT as a diagnostic, we characterize the critical behaviors with both the quantum metric and Berry curvature, which quantifies the response of the system to a variation of the governing Hamiltonian in the Hilbert space, and obtain a ground-state phase diagram between normal and superradiation phases. The numerical results demonstrate that the quantum metric and Berry curvature are dramatically changed even for a tiny variation of the control parameter near the critical point, serving as good signatures for the QPT. Notably, the location of the quantum critical point remains unchanged with an increase in the two-photon driving phase $\phi$. The finite-size scaling of the geometric tensor, critical exponent, and scaling dimension is obtained by an exact numerical method, which is consistent with previous works \cite{RN26}. Our numerical results pinpoint that the phase transition falls within the universality class of the quantum Rabi model. The revealed divergent behaviors and universal scaling shed new light on the critical phenomena of the nonlinear resonator. Interesting future questions include the exotic phase and phase transitions in parametrical-driven Kerr resonators with dissipation.

\begin{acknowledgements}
This work was supported by the National Natural Science Foundation of China (Grants No. 12274080, No. 11875108), National Youth Science Foundation of China (No. 12204105), the Educational Research Project for Young and Middle-aged Teachers of Fujian Province (Grant No. JAT210041) and the Natural Science Foundation of Fujian Province (Grants No. 2022J05116). X.-J.Yu acknowledges support from the start-up grant XRC-23102 of Fuzhou University.
\end{acknowledgements}

\appendix
\section{Equivalence of the quantum metric and the fidelity susceptibility}
\label{appendixA}

In this section, we provide detailed proof of the equivalence between the quantum metric tensor and the fidelity susceptibility. The Hamiltonian in Eq. (\ref{Hd}) can be expressed as follows:
\begin{equation}
H(\varepsilon) = H_0 + \varepsilon H_1,
\end{equation}
where $H_0$ is the Hamiltonian of the nonlinear oscillator and $H_1$ is the perturbation inducing QPTs from the normal to the superradiant phase with the driving parameter $\varepsilon$. The quantum ground-state fidelity $F(\varepsilon, \varepsilon + \delta_\varepsilon)$, defined as the overlapping amplitude of the ground state wavefunction with the driving parameter $\varepsilon$ and the ground state wavefunction with the driving parameter $\varepsilon + \delta_\varepsilon$, is given by:
\begin{equation}
F(\varepsilon, \varepsilon + \delta_\varepsilon) = \vert \bra{u_0(\varepsilon)}\ket{u_0(\varepsilon + \delta_\varepsilon)} \vert.
\end{equation}
Its value is almost zero near $\varepsilon_{c}^{*}$, i.e., $F(\varepsilon_{c}^{*}, \varepsilon_{c}^{*} + \delta_\varepsilon) \sim 0$. The ground states exhibit substantial differences on each side of a quantum critical point, leading to a pronounced dip in the vicinity of this critical point.

In practice, a more convenient quantity to characterize QPTs is the fidelity susceptibility, defined by the leading term of the fidelity~\cite{RN44}:
\begin{equation}
\chi_F(\varepsilon) = \lim_{\delta_\varepsilon\rightarrow 0}\frac{-2\ln F}{(\delta_\varepsilon)^2}.
\end{equation}
To directly show the equivalence between the two physical quantities, expressed in terms of the eigenstates of the Hamiltonian, the fidelity susceptibility is represented as:
\begin{equation}
\chi_F(\varepsilon) = \sum_{n=1}\frac{\vert\bra{u_{\rm n}(\varepsilon)}H_1\ket{u_0(\varepsilon)} \vert^2}{[E_{\rm n}(\varepsilon)-E_0(\varepsilon)]^2},
\end{equation}
where $\ket{u_{\rm n}(\varepsilon)}$ is the eigenstate of $H(\varepsilon)$ with the corresponding eigenenergy $E_{\rm n}(\varepsilon)$. Additionally, we observe that $H_1=\partial_\varepsilon H$. Therefore, by comparing the expression of Eq. (\ref{QmunuH}), we have proven that the fidelity susceptibility is indeed the quantum metric tensor, i.e., $\Re{Q_{\varepsilon\varepsilon}}=g_{\varepsilon\varepsilon}$.

\bibliography{reference}
\end{document}